\documentclass[9pt]{article}

\usepackage{url}
\usepackage{caption}
\usepackage[usenames,dvipsnames]{xcolor}
\usepackage{ragged2e}
\usepackage[cmex10]{amsmath}
\usepackage{float}
\usepackage{graphicx}
\usepackage{amssymb,amsfonts}
\usepackage{fullpage}
\usepackage{times}
\usepackage{overpic}
\usepackage{nameref,hyperref}

\graphicspath{{figures/}}
\DeclareGraphicsExtensions{.pdf,.png,.jpg}

\newcommand{\reffig}[1]{Figure\,\ref{#1}}
\newcommand{\refeq}[1]{equation\,(\ref{#1})}
\newcommand{\reftab}[1]{Table\,\ref{#1}}

\newcommand{\jni}{\justifying\noindent}

\makeatletter
\renewcommand\@biblabel[1]{#1 }

\def\@maketitle{%
  \newpage\spacing{1}\setlength{\parskip}{10pt}%
    \begin{center}%
	{\large\bfseries\noindent\sloppy \textsf{\@title} \par}%
    {\noindent\sloppy \@author}%
	\end{center}%
}

\renewcommand{\section}{\@startsection {section}{1}{0pt}%
    {-6pt}{1pt}%
    {\reset@font \large \bfseries}%
    }
\renewcommand{\subsection}{\@startsection {subsection}{2}{0pt}%
    {-6pt}{1pt}%
    {\reset@font \normalsize \bfseries}%
    }
\makeatother

\newcommand{\spacing}[1]{\renewcommand{\baselinestretch}{#1}\large\normalsize}

\AtBeginDocument{}

\begin{document}
\begin{flushleft}
{\Large
\textbf\newline{Hybrid spreading mechanisms and T cell activation shape the dynamics of HIV-1 infection}
}
\newline
\\
Changwang Zhang\textsuperscript{1,2,4}, 
Shi Zhou\textsuperscript{1}, 
Elisabetta Groppelli\textsuperscript{3},
Pierre Pellegrino\textsuperscript{5},
Ian Williams\textsuperscript{5},
Persephone Borrow\textsuperscript{6},
Benjamin M. Chain\textsuperscript{3,\ddag,*},
Clare Jolly\textsuperscript{3,\ddag,*}	
\\
\bf{1} Department of Computer Science, University College London, London, United Kingdom
\\
\bf{2} Security Science Doctoral Research Training Centre, University College London, London, United Kingdom
\\
\bf{3} Division of Infection and Immunity, University College London, London, United Kingdom
\\
\bf{4} School of Computer Science, National University of Defense Technology, Changsha, China
\\
\bf{5} Centre for Sexual Health \& HIV Research, Mortimer Market Centre, London, United Kingdom
\\
\bf{6} Nuffield Department of Medicine, University of Oxford, Oxford, United Kingdom
\\
\ddag These authors are joint last authors on this work.
\\
* E-mail: b.chain@ucl.ac.uk (BC); c.jolly@ucl.ac.uk (CJ)
\\
\end{flushleft}

\section*{Abstract}
HIV-1 can disseminate between susceptible cells by two mechanisms: cell-free infection following fluid-phase diffusion of virions and by highly-efficient direct cell-to-cell transmission at immune cell contacts. The contribution of this hybrid spreading mechanism, which is also a characteristic of some important computer worm outbreaks, to HIV-1 progression in vivo remains unknown. Here we present a new mathematical model that explicitly incorporates the ability of HIV-1 to use hybrid spreading mechanisms and evaluate the consequences for HIV-1 pathogenenesis. The model captures the major phases of the HIV-1 infection course of a cohort of treatment naive patients and also accurately predicts the results of the Short Pulse Anti-Retroviral Therapy at Seroconversion (SPARTAC) trial. Using this model we find that hybrid spreading is critical to seed and establish infection, and that cell-to-cell spread and increased CD4$^+$ T cell activation are important for HIV-1 progression. Notably, the model predicts that cell-to-cell spread becomes increasingly effective as infection progresses and thus may present a considerable treatment barrier. Deriving predictions of various treatments' influence on HIV-1 progression highlights the importance of earlier intervention and suggests that treatments effectively targeting cell-to-cell HIV-1 spread can delay progression to AIDS. This study suggests that hybrid spreading is a fundamental feature of HIV infection, and provides the mathematical framework incorporating this feature with which to evaluate future therapeutic strategies.


\section*{Author Summary}
The ability to spread using more than once mechanism, named hybrid spreading, is a ubiquitous feature of many real world epidemics including HIV and Hepatitis C virus infection \textit{in vivo}, and computer worms spreading on the Internet. Hybrid spreading of HIV is well documented experimentally but its importance to HIV progression has been unclear. In this paper, we introduce a mathematical model of HIV dynamics that explicitly incorporates hybrid spreading. The model output shows excellent agreement to two sets of clinical data from a treatment naive cohort and from the Short Pulse Anti-Retroviral Therapy at Seroconversion trial. The model demonstrates that hybrid spreading is an essential feature of HIV progression, a result which has significant implications for future therapeutic strategies against HIV.

\section*{Introduction}

The course of HIV-1 infection is typified by three phases; acute infection characterized by a rapid viraemia peak (3-6 weeks post-infection) followed by a rapid fall in virus levels, a stable chronic phase of variable length characterized by low level viraemia and slowly declining CD4$^+$ T cell numbers, and a final stage (Acquired Immune Deficiency Syndrome, AIDS) characterized by multiple opportunistic infections and a rapid fall in CD4$^+$ T cell count. The cellular and viral changes which drive each phase of this complex infection have been the subject of intense debate, in which mathematical models have played an important role in delineating HIV-1 pathogenesis and informing antiretroviral therapy \cite{R6_Perelson_2013,R7_Ho_1995,R8_Perelson_1996}. The rich literature appertaining to mathematical modeling of intra host HIV dynamics has been reviewed several times recently  \cite{R6_Perelson_2013,Prague_2013,Xiao_2013,R5_Alizon_2012}.  Recent studies incorporate sophisticated models of immune selection \cite{Elemans_2014,Deutekom_Boer_2013}, as well as the formation of a latent reservoir of quiescent infected cells \cite{Rong_Perelson_2009,Spina_2013}. However the interplay of cell-to-cell spread and increased CD4$^+$ T cell activation, that are likely to have profound influences on the progression of the disease have been hitherto little studied. Here we have addressed this and developed a unified model that can explain the complex progress of the infection in all its phases and its variable timescale. Such a unified model is important not only to understand the HIV-1 infection dynamics, but also to estimate the long term effects of therapeutic strategies on HIV-1 progression. In this paper, unless otherwise stated, ``T cells'' refers to CD4$^+$ T cells.

HIV-1 predominantly replicates in CD4$^+$ T cells \textit{in vivo}, and is now known to spread between T cells by two parallel routes. According to the classical model of HIV-1 spread, virus particles bud from an infected T cell, enter the blood/extracellular fluid and then infect another T cell following a chance encounter (termed cell-free spread). Because diffusion of virus particles is much faster than cell migration, and there is extensive flow of blood and fluid, this mode of spreading can be characterized by a well mixed epidemic spreading model. In this scenario, the probability of infection for a particular cell will be proportional to the concentration of extracellular infectious virus. However, HIV-1 can also disseminate by direct transmission from one cell to another by a process of cell-to-cell spread. Two pathways of cell-to-cell transmission have been reported. Firstly, an infected T cell can transmit virus directly to a target T cell via a virological synapse \cite{R1_Sattentau_2008,R13_Jolly_2004,Komarova_2013_SR}. Secondly, an antigen presenting cell (APC) can also transmit HIV-1 to T cells by a process that either involves productive infection (in the case of macrophages) or capture and transfer of virions \textit{in trans} (in the case of dendritic cells)  \cite{R1_Sattentau_2008}. Whichever pathway is used, infection by cell-to-cell transfer is reported to be much more efficient than cell-free virus spread \cite{R14_Martin_2010,R15_Sourisseau_2007,Duncan_2013}. A number of factors contribute to this increased efficiency, including polarised virus budding towards the site of cell-to-cell contact, close apposition of cells which minimizes fluid-phase diffusion of virions, and clustering of HIV-1 entry receptors on the target cell to the contact zone \cite{R1_Sattentau_2008, R13_Jolly_2004}. Cell-to-cell spread is thought to be particularly important in lymphoid tissues where CD4$^+$ T lymphocytes are densely packed and likely to frequently interact. Indeed, intravital imaging studies have supported the concept of the HIV-1 virological synapse \textit{in vivo} \cite{R25_Murooka_2012,Sewald_2012}.

Hybrid spreading is in fact a feature of other viral infections \cite{Zeisel_2013}, but is also shared in other ``epidemic'' scenarios such as spread of computer worms \cite{Moore_2002,Zhang_2014}, or of mobile phone viruses \cite{Wang_2009}. The mathematical analysis of hybrid spreading has received significant previous attention \cite{Wang_2009, Keeling_2010,Balcan_2011, Zhang_hm_2014}. However, the importance of hybrid spread to HIV-1 dissemination and disease progression, has not been explored from a mathematical point of view.


In this paper we develop a new mathematical model which incorporates the basic principles of previous host-centric models including a virus-dependent immune response \cite{Deutekom_Boer_2013}, viral latency and a progressive increase in cell activation \cite{R2_Miedema_2013,R24_Doitsh_2014}. Notably, the model additionally includes explicit terms for the two modes of virus spread, parametrised from experimental observation. The model faithfully replicates the overall three phase course of HIV-1 infection . The model predictions are consistent with both a set of longitudinal data (viral load and CD4$^+$ T cell count) from a cohort of treatment naive HIV-1 infected patients  and the results of the Short Pulse Anti-Retroviral Therapy at Seroconversion (SPARTAC) trial that aims to evaluate how the short-course antiretroviral therapy (ART) delays HIV progression \cite{SPARTAC_2013}. The results of our study reveal the importance of two modes of HIV-1 spread, highlight the close link between cell-to-cell spread and cell activation in driving the progression of HIV-1 infection to AIDS and support early therapeutic intervention (i.e. ``test-and-treat'' initiatives) to delay disease progression in infected individuals. Since cell-to-cell spread is likely to present a considerable barrier to HIV-1 eradication, our data suggest that efforts to target this mode of viral spread whilst simultaneously manipulating CD4$^+$ T cell activation may be a fruitful strategy to help control virus infection and halt progression to AIDS.

\section*{Results}

\subsection*{The HIV-1 model}

\begin{figure}[h]\centering\small
\begin{overpic}[width=0.5\textwidth]{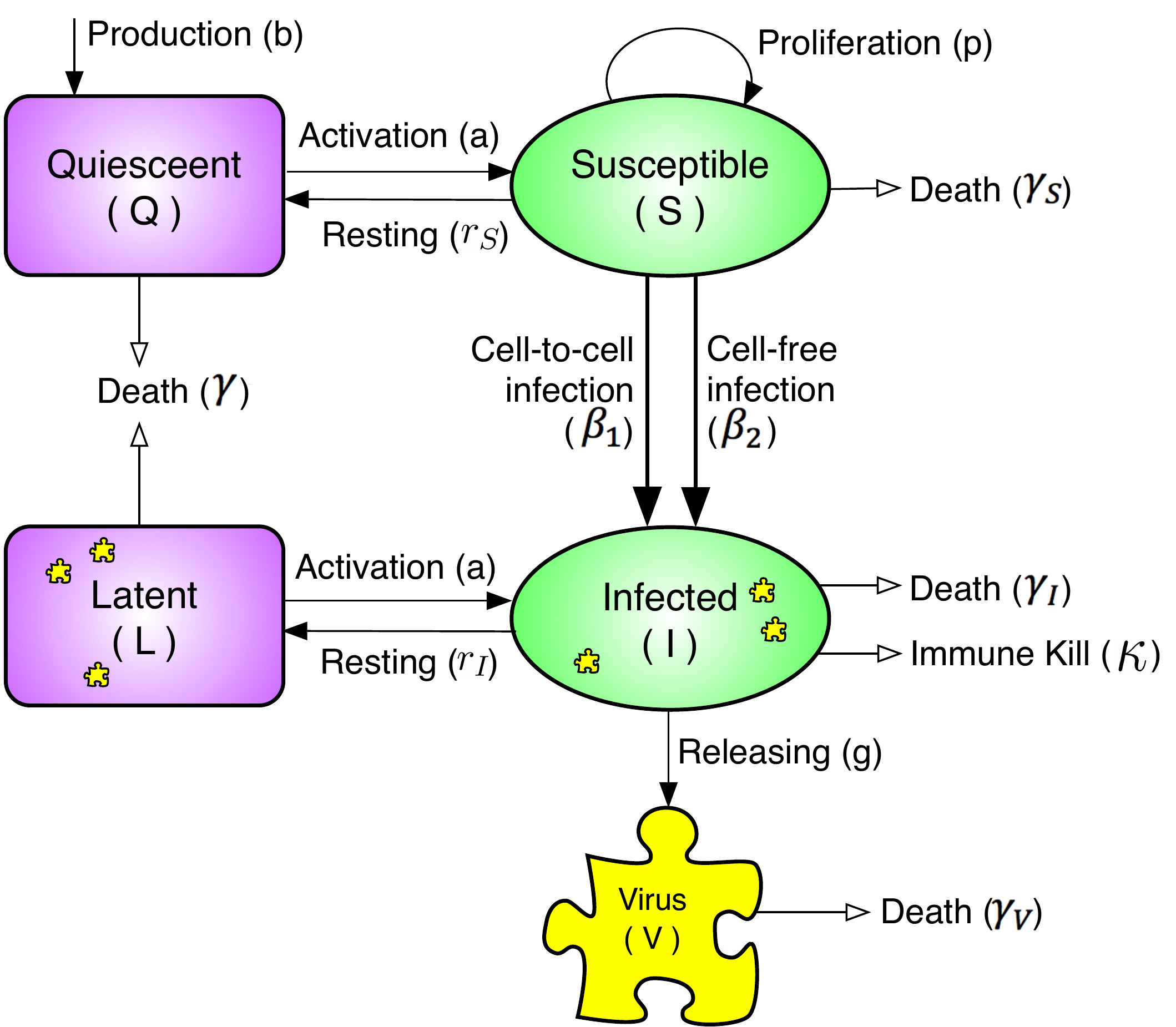}
\put(0,82){\textbf{A}}
\end{overpic}\\
\begin{overpic}[width=0.49\textwidth]{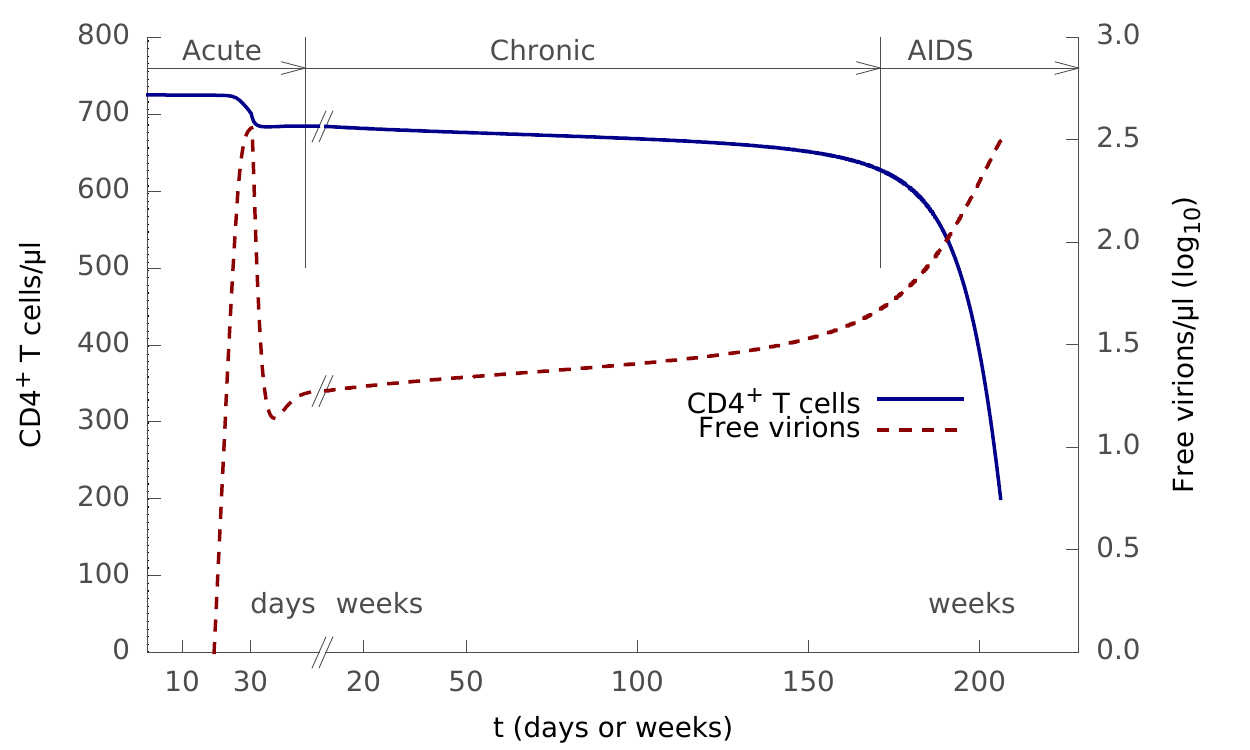}
\put(0,50){\textbf{B}}
\end{overpic}
\begin{overpic}[width=0.49\textwidth]{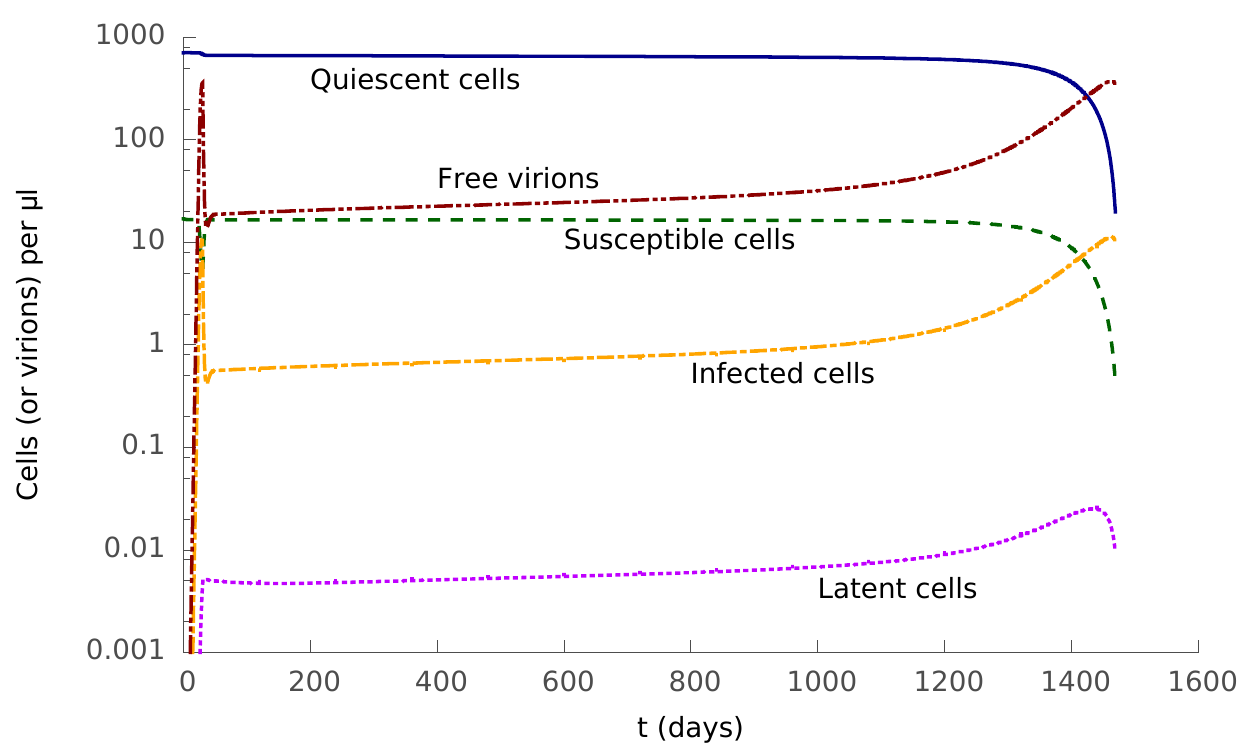}
\put(0,50){\textbf{C}}
\end{overpic}
\caption{\label{fig-mod}\textbf{The HIV-1 model  reproduces the full course of HIV-1 infection.} \textbf{(A)} Diagrammatic representation of the model described by \refeq{eq:hiv-model2}.  \textbf{(B)} Numerical solutions of the model,  plotting $N$, the density of all CD4$^+$ T cells (y axis on the left)
and $V$, the density of free virions (y axis on the right in log scale) as a function of time (in days or weeks), respectively. The initial infection starts on day 0 and the cellular immune response starts on day 30. Parameter values are in \reftab{tab-pars}.\textbf{(C)} The density of quiescent, susceptible, latent and infected CD4$^+$ T cells, and the density of free virions as function of time (in days).}
\end{figure}

We here introduce a model of HIV-1 infection as depicted in \reffig{fig-mod}A. We consider four distinct CD4$^+$ T cell states: activated, uninfected susceptible (\textit{S}) cells; activated and productively infected (\textit{I}) cells ; quiescent, uninfected (\textit{Q}) cells; and quiescent latently (\textit{L}) infected cells. The total CD4$^+$ cell density (N) changes with time and is given by the sum of these four terms, i.e. $N(t)=Q(t)+S(t)+I(t)+L(t)$. The model can be described by an Ordinary Differential Equation (ODE) system (\refeq{eq:hiv-model2}) and is illustrated in \reffig{fig-mod}A.
\begin{equation}
\begin{aligned}\label{eq:hiv-model2}
&\frac{dQ}{dt}=-\gamma Q + r_S S -a\dfrac{N_{M}}{N}Q +b\\
&\frac{dS}{dt}=-\gamma_SS -r_S S +a\dfrac{N_M}{N}Q -cI\dfrac{S}{N}\theta\beta_1-SV\beta_2+p\dfrac{N_M-N}{N_M}S\\
&\frac{dI}{dt}= -\gamma_II - r_I I +a\dfrac{N_{M}}{N}L  +cI\dfrac{S}{N}\theta\beta_1+SV\beta_2-\kappa \frac{I}{I+0.1}\frac{N}{N_M} I \\
&\frac{dL}{dt}=-\gamma L +r_I I-a\dfrac{N_{M}}{N}L\\
&\frac{dV}{dt}=-\gamma_VV +gI
\end{aligned}
\end{equation}
The density variables (\textit{Q, S, I, L, V}) and parameters are defined in \reftab{tab-pars}. The densities are measured as numbers of cells or virions in a $\mu l$ of blood/extracellular fluid. We set a density variable to zero when it drops to below $10^{-12}/\mu l$, accounting for the fact that when density of cells or virions drops to such low level, there is a high probability that it would die out (density becomes zero). The default value of parameters, shown in \reftab{tab-pars}, are taken from the literature or estimated from clinical and experimentally observed data. The killing coefficient $\kappa$ equals to its value in Table 1 when $t \geq D$; otherwise $\kappa=0$ when $t<D$.

\begin{table}[h]\small\centering
\caption{\label{tab-pars}\textbf{Variables and parameters of our model }}
\begin{tabular}{clrll}
\hline
& {\bf Variables} & Initial value & Unit & \\ 
\hline
$Q$ & Density of quiescent CD4$^+$ T cells $^a$  & $Q_0=708$  & $cells/\mu l$ \\
$S$ & Density of susceptible CD4$^+$ T cells $^a$  & $S_0=17$  & $cells/\mu l$\\
$I$ & Density of infected CD4$^+$ T cells & $I_0=0$ & $cells/\mu l$\\
$L$ & Density of latent CD4$^+$ T cells & $L_0=0$ & $cells/\mu l$\\
$N$& Density of all CD4$^+$ T cells   & $N_0=725$ & $cells/\mu l$\\
$V$ & Density of free HIV-1 virus & $V_0=10^{-6}$ & $virions/\mu l$\\
\hline
& {\bf Parameters} & Default value  & Unit & Reference\\ 
\hline

$\gamma$ & Death rate of  quiescent and latent cells & $0.001$ & $/day$ &\cite{R12_De_1998}\\
$\gamma_S$ & Death rate of susceptible cells & $0.0625$ & $/day$ &\cite{R46_Mohri_2001}\\
$\gamma_I$ & Death rate of infected cells  & $0.5$ & $/day$ &\cite{R12_De_1998}\\
$b$ & Production rate of new quiescent  cells $^b$   & $0.17$ & $cells/\mu l /day$ &\cite{R40_Stafford_2000}\\ 
$r_S$ & Resting rate of susceptible cells   & $0.5$ & $/day$ \\
$r_I$ & Resting rate of infected cells   & $0.0001$ & $/day$\\

$a$ & Cell activation coefficient & $0.01$ & $/day$\\
$p$ & Cell proliferation coefficient & $1$ & $/day$ &\cite{R12_De_1998}\\
$N_M$ & Cell threshold density   $^c$ & $800$ & $cells/\mu l$ \\
\hline
$g$ & Virus release rate   from an infected cell & $100$ & $virions/cell/day$ &\cite{R12_De_1998}\\
$\gamma_V$ & Virus death rate   & $3$ & $/day$ &\cite{R40_Stafford_2000,R12_De_1998} \\
\hline
$c$ & Cell contact rate $^d$ & $330$ & $/day$ & \cite{R14_Martin_2010,R41_Jolly_2007,R43_Jolly_2010}\\
$\theta$ & Cell synapse rate $^e$ & $0.56$ &&\cite{R13_Jolly_2004,R14_Martin_2010,R41_Jolly_2007,R42_McNerney_2009}\\
$\beta_1$ & Cell-to-cell infection rate $^f$ & $0.19$ &&\cite{R15_Sourisseau_2007,R43_Jolly_2010,R44_Chen_2007,R45_Del_2011}\\
$\beta_2$ & Cell-free infection rate & $0.00135$ & $\mu l/virions/day$ &\cite{R12_De_1998}\\
\hline
$D$ & Immune response delay  $^g$ & $30$ & $days$&\cite{R47_McMichael_2010}\\
$\kappa$& Killing coefficient (when $t\geqslant D$) $^h$  & $1.5$ & $/day$\\
\hline
\end{tabular}\\
\jni $^a$ $Q_0$ and $S_0$ are chosen to maintain a stable density of CD4$^+$ T cells ($N_0=725 \, cell/\mu l$) in the absence of an infection; 
$^b$ Production from sources including thymus;
$^c$ Beyond this threshold CD4$^+$ T cell proliferation stops;
$^d$ The average number of effective contacts between two CD4$^+$ T cells. Cells contact each other frequently, but only a proportion ($20-25\%$) of these random contacts are sufficiently stable and ``effective''  for a synapse to be formed at rate $\theta$;
$^e$ The average probability to form a virological synapse when two cells have an effective contact;
$^f$ The average probability that a susceptible cell is infected when it forms a synapse with an infected cell;
$^g$ Between the initial infection and the onset of cellular immune response, including the period before  symptoms begin;
$^h$ $\kappa=0$ when $t<D$.
\end{table}

The production rate of new quiescent T cells from sources, such as thymus, within the human body is represented by $b$. Quiescent T cells are activated and become susceptible at a variable activation rate $a(N_M/N)$, where $a$ is the activation coefficient and $N_M$ is the density of T cells  at which proliferation stops. The activation rate $a(N_M/N)$ increases as the total T cell density ($N$) falls (caused by the HIV-1 progression). The detailed mechanism of how the activation rate increases with the progression of HIV-1 is still unclear, and may include increased rates of co-infections, danger signals from dying T cells or homeostatic regulatory loops. This term, $a(N_M/N)$, here is an approximation, which encompasses the combined effects of all these different mechanisms. Quiescent T cells die at a rate of $\gamma$.

Susceptible T cells turn into quiescent T cells at a rate $r$. They proliferate at a variable rate $p(1-N/N_M)$, where $p$ is the proliferation coefficient, $N$ is the total T cell density, and $N_M$ is the T cell density at which proliferation stops. This variable proliferation rate is a reasonable approximation \cite{R12_De_1998} to the real T cell proliferation process, based on evidence \cite{R7_Ho_1995} that T cell proliferation rate is density-dependent and would slow as the T cell density becomes high. Susceptible T cells die at a rate $\gamma_S$.

Susceptible T cells become infected through both cell-to-cell and cell-free infection. For cell-free infection, the newly infected susceptible T cells per unit period of time are $\beta_2SV$, where $\beta_2$ is the infection rate of susceptible T cells by free virus. For cell-to-cell infection, we consider that the T cells are randomly moving, i.e. a T cell has an equal chance of contacting any other T cells. Let $c$ represent the number of effective contacts each T cell makes in a unit period of time. Then the number of contacts made by all infected T cells per unit period of time is $c I$. Among those contacts, $S/N$ are contacts with a susceptible T cell that could potentially end up with a new infection. Let $\theta$ represent the Synapse rate: the average probability that two T cells form a virological synapse once they have made an effective contact. Then the newly infected susceptible T cells through cell-to-cell infection per unit time period can be represented as $cI(S/N)\theta\beta_1$, where $\beta_1$ is the cell-to-cell infection rate when an infected T cell and a susceptible T cell form a synapse. In reality, cell-to-cell transfer occurs locally involving only the infected cell and its immediate neighbours. The model abstracts this process by averaging infection over all cells. In practice, local effects will only distort this average when target cells in the vicinity of an infected cell become limiting. This limit seems unlikely to be reached except very late in infection, given that infected cells continue to migrate, albeit at a slower rate (personal observations and \cite{R25_Murooka_2012}), and uninfected target cells continue to migrate into the vicinity of an infected cell. More complex spatial models will be required, however, to understand the detailed anatomical distribution of HIV-1 infected cells over time. The amount of cell-to-cell transfer in the model depends on the number of infected T cells $I$ and the proportion of susceptible cells $S/N$ at any given time. Since $N$, the total density of CD4$^+$ cells,  is not held constant, but in fact declines over time, cell-to-cell spread becomes increasingly effective as HIV-1 progresses.

The model does not explicitly distinguish antigen-presenting cell  to T cell (APC/T) from T cell to T cell (T/T) transmission. APC/T  transmission may potentially be most important very early in establishing infection \cite{R26_Parrish_2013}, a process which is not examined in detail in this model. These two types of transmission are in fact both likely to occur most frequently and efficiently in the microenvironment of an APC/T  cluster, where APC/T  interactions lead to T cell activation, and hence favour also T/T  interaction. Furthermore, there are very few quantitative estimates of the parameters of APC/T     interaction in vivo. The incorporation of an additional cell type is therefore unlikely to have a major effect on the model behaviour, but would add significantly to model complexity and uncertainty.

The cause of  T cell death in HIV-1 infection continues to be controversial, and probably includes several effects including lysis of infected cells by effector cells such as CD8 T cells and NK cells, apoptosis/pyroptosis and bystander death \cite{R24_Doitsh_2014}.  Non-immunological death of infected T cells is represented by a death rate of $\gamma_I$. And we use the term $\kappa \frac{I}{I+0.1} \frac{N}{N_M}$ to model the death of infected T cells by the cellular immune response. $\kappa$ is initially $0$ and changes to a higher value in \reftab{tab-pars} when the cellular immune response kicks in $D$ (default value: 30) days after the initial infection. The term $\frac{I}{I+0.1}$ captures the relationship between the strength of the immune response and the density of infected CD4$^+$ T cells \cite{Deutekom_Boer_2013}. The term $N/N_M$ captures immune exhaustion caused by HIV-1 infection. It falls from around 1 before infection towards 0 as HIV-1 progresses (because $N_M$ is a constant and $N$, the total density of CD4$^+$ T cells, gradually declines with the HIV-1 progression).

Infected cells return to a quiescent state, and become latent, at rate $r_I$. Latent cells die and are activated (i.e. becomes infected cells) at the same rates as quiescent cells. Infected T cells release free viruses at rate $g$. Free viruses die at a rate of $\gamma_V$.

The abortive infection of quiescent cells is not considered in this simplified model, similarly to most previous modelling studies \cite{R8_Perelson_1996,R6_Perelson_2013,R19_Komarova_2013,R20_Komarova_2013}. HIV-1 immune escape mutants \cite{Fryer_2010} are not directly modelled in this paper but their effects on degrading cellular immune response are reflected in the immune exhaustion in our model.

The numerical solutions of each of the variables are shown in \reffig{fig-mod}C. \reffig{fig-mod}B shows the combined CD4$^+$ T cell counts ($N$) and virus load ($V$), which are measured routinely in the clinic to monitor HIV-1 infection. Notably, the qualitative behaviour of the model accurately reflects the three main phases of disease that are observed clinically. The model reproduces an acute infection phase, where the virus replicates rapidly (reflecting the absence of any pre-existing adaptive immunity), peaks and then returns to a low level by approximately 5 weeks. This metastable level of virus represents the clinical ``set-point''.  Virus then remains stable for a prolonged period (note interruption and change of scale in x axis), during which time T cells decline very slowly. Finally, T cell numbers start to drop faster, and viral loads rise. The model calculation is stopped when CD4$^+$ level reaches $200\,cells/\mu l$.

It is important to note that the term $cI(S/N)\theta\beta_1$ we introduced for cell-to-cell spread is mathematically different from the classic term $\beta_2SV$ which captures cell-free spread. $N$, the total CD4$^+$ number, is not a constant but decreases as a function of time. Indeed, if $N$ in the cell-to-cell spread term is fixed, disease progression is not observed in the model (\nameref{fig-constn}). Furthermore, $N$ cannot be expressed as a simple function of $V$. In \reffig{fig-mod}B the same value of $V$ (free virion density) on each side of its peak around $t=30$ days corresponds to two different $N$ (CD4$^+$ T cell density) values. In addition, I, the number of infected cells, is not always proportional to V, especially early in infection (\nameref{fig-vdi}). Thus the term for cell-to-cell spread $cI(S/N)\theta\beta_1$ shows different dynamics from the classic cell-free term $\beta_2SV$.

\subsection*{Model Evaluation}

Most of the parameters of our model, and especially those determining HIV infectivity via cell-to-cell or cell-free spreading, were obtained from experimental observations (\reftab{tab-pars}). It was therefore important to test that the model with this parametrization accurately fits real clinical data sets. We therefore evaluated our model against a set of longitudinal T cell count and virus load measurements obtained from a cohort of HIV-1-infected individuals who were recruited following clinical presentation with symptomatic acute HIV-1 infection and followed over time with serial measurement of plasma viral RNA levels and circulating T cell counts. The subjects selected for inclusion in this study all chose not to receive antiretroviral treatment in acute or early infection, and remained untreated until progression towards AIDS, evidenced by a substantial decline in their circulating CD4$^+$ T cell count (See Materials and Methods). 

\begin{figure}[H]\centering\small
\begin{overpic}[width=0.6\textwidth]{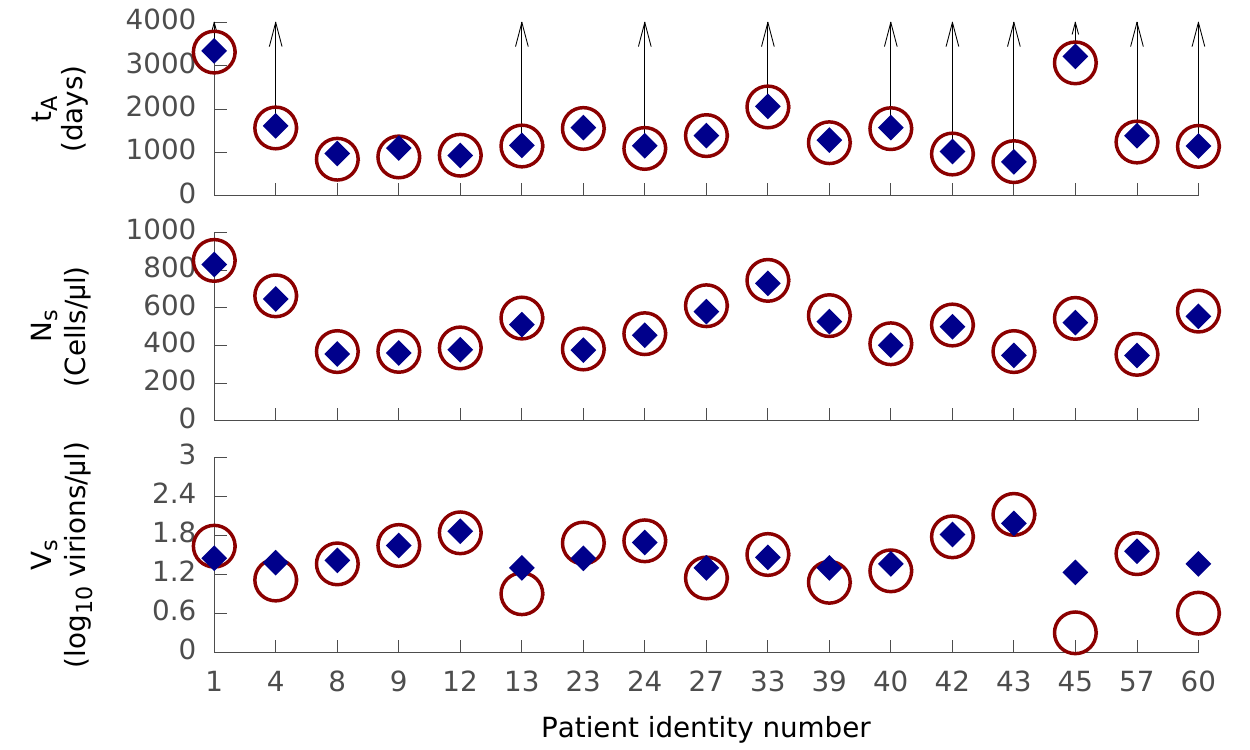}
\put(4,56){\textbf{A}}
\end{overpic}\\
\begin{overpic}[width=0.49\textwidth]{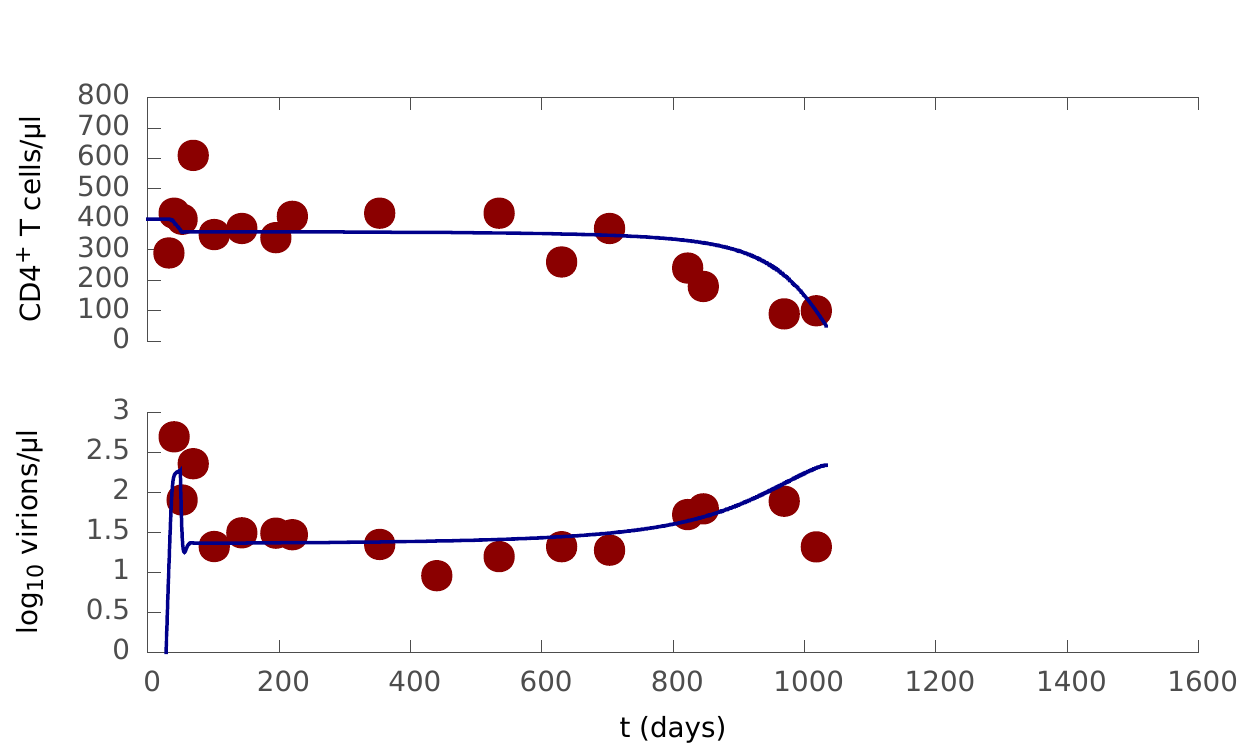}
\put(1,55){\textbf{B}}
\put(40,30){\textbf{Patient MM8}}
\end{overpic}
\begin{overpic}[width=0.49\textwidth]{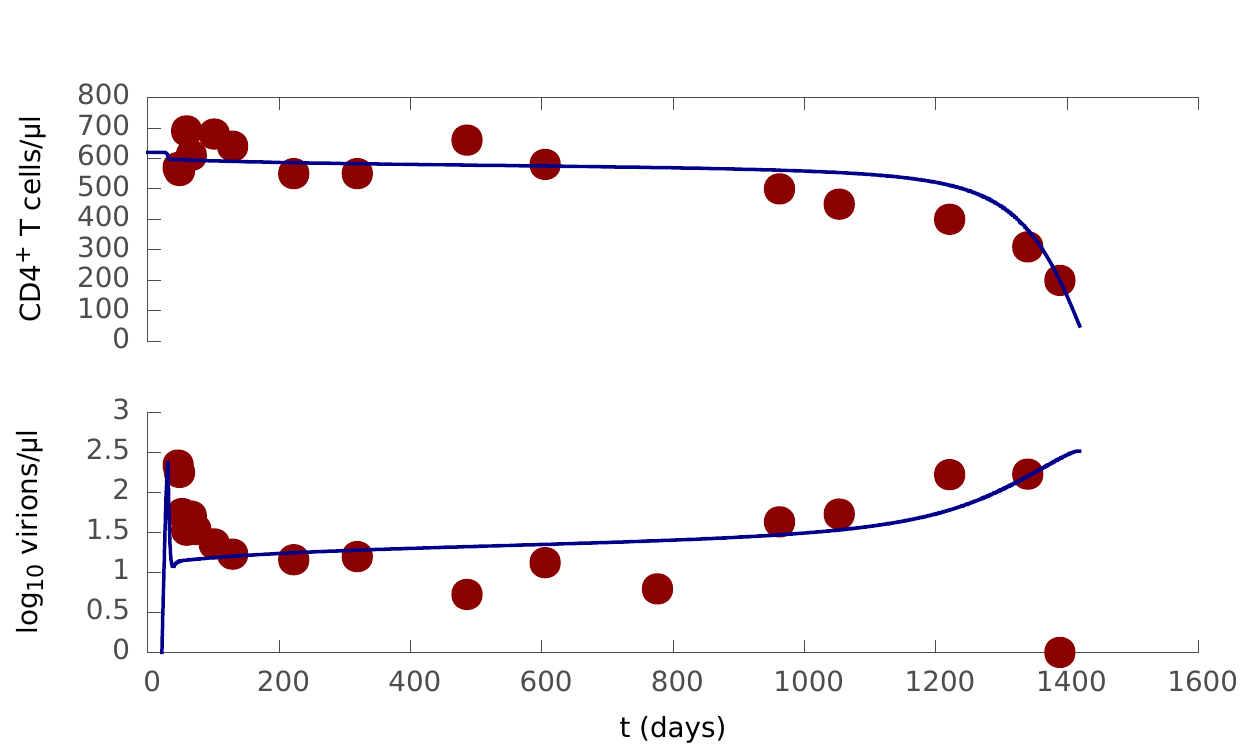}
\put(40,30){\textbf{Patient MM27}}
\end{overpic}
\begin{overpic}[width=0.49\textwidth]{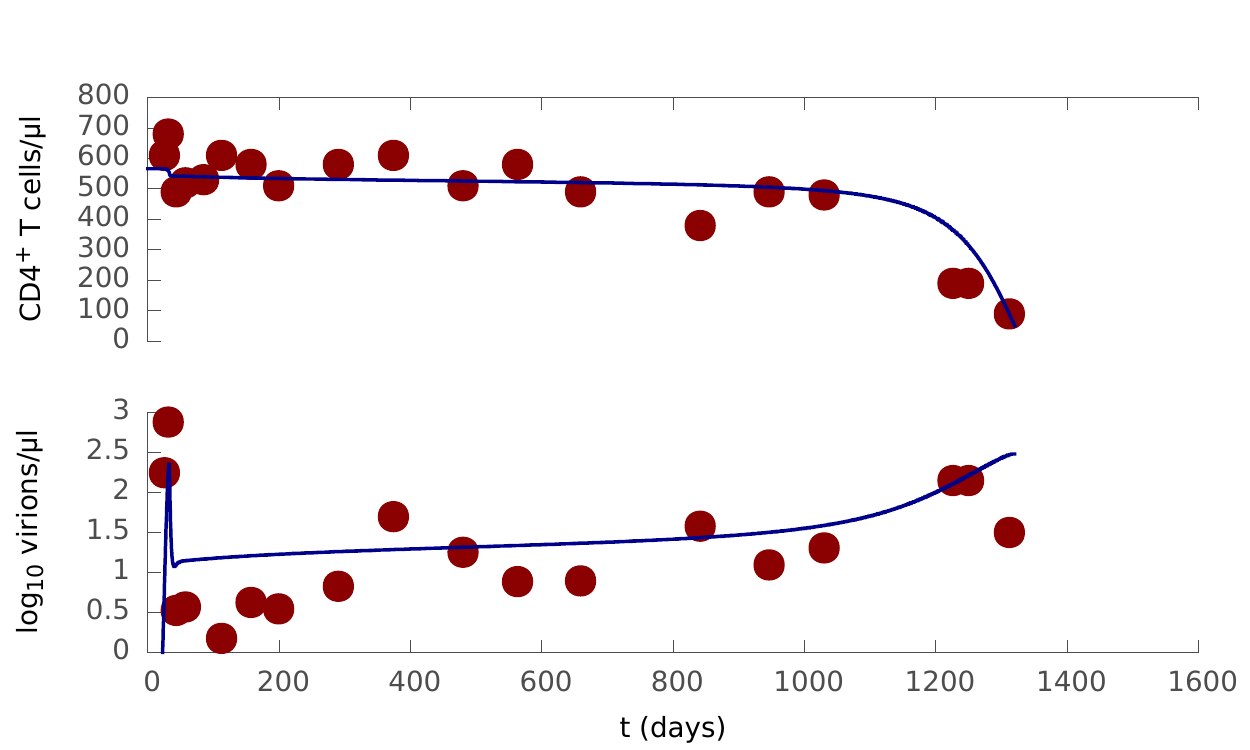}
\put(40,30){\textbf{Patient MM39}}
\end{overpic}
\begin{overpic}[width=0.49\textwidth]{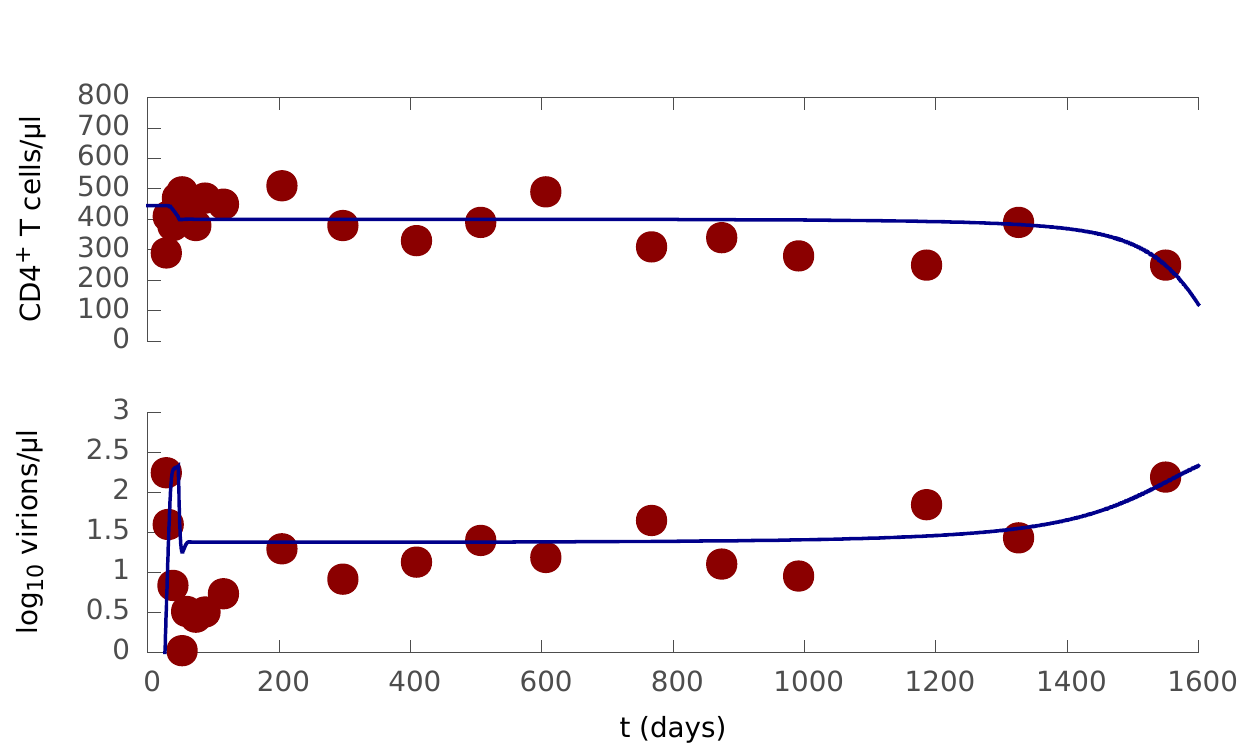}
\put(40,30){\textbf{Patient MM40}}
\end{overpic}
\caption{\label{fig-fit}\textbf{Model prediction for a cohort of treatment-naive HIV-1 patients.}\textbf{(A)} Clinical data (circle and arrow) for all patients under study comparing against model prediction (diamond) for the time to AIDS ($t_A$), the quasi-steady density of  CD4$^+$ T cells ($N_s$) and the  quasi-steady density of free virions ($V_s$). An arrow represents that $t_A$ is greater than a particular value (represented by the connected circle) for a patient as his / her CD4$^+$ count did not reached AIDS level ($200\,cells/\mu l$) in the data.\textbf{(B)} Prediction (curve) of HIV progression course ($N$ and $log_{10}V$) for four typical patients, where clinical data are shown as dots.Full prediction results are shown in \nameref{tab-fit-pars} and \nameref{tab-fit-vars}.}
\end{figure}

We use our model to theoretically reproduce the HIV-1 infection courses in the data. The values for HIV infectivity were fixed, as derived from the literature or our own observations (\reftab{tab-pars}). Values for five parameters ($Q_0$, $S_0$, $N_M$, $\kappa$ and $D$) describing the characteristics of the immune response were chosen for each patient to minimise the error of the predicted quasi-stable level of  T cell counts ($N_s$) and viral load ($V_s$), and  the time of progression to AIDS ($t_A$). All other values remained fixed at the default values in \reftab{tab-pars}. The predicted progression results are compared against the actual measurements in \reffig{fig-fit} and \nameref{tab-fit-vars}. The predicted $V_s$ and $t_A$ for each patient were negatively correlated (correlation coefficient = $-0.46$), in agreement with the well-established relationship between these two clinical values. Remarkably, the model can fit all patients by modifying the five immune-relevant parameters over a narrow range. Furthermore, the parameter values which gave the best results for the patients (see \nameref{tab-fit-pars}) are all very close to those in \reftab{tab-pars}, which were derived independently from experimental measurements.

\subsection*{The importance of cell-to-cell spread and cellular activation}

\begin{figure}[h]\centering\small
\begin{overpic}[width=0.49\textwidth]{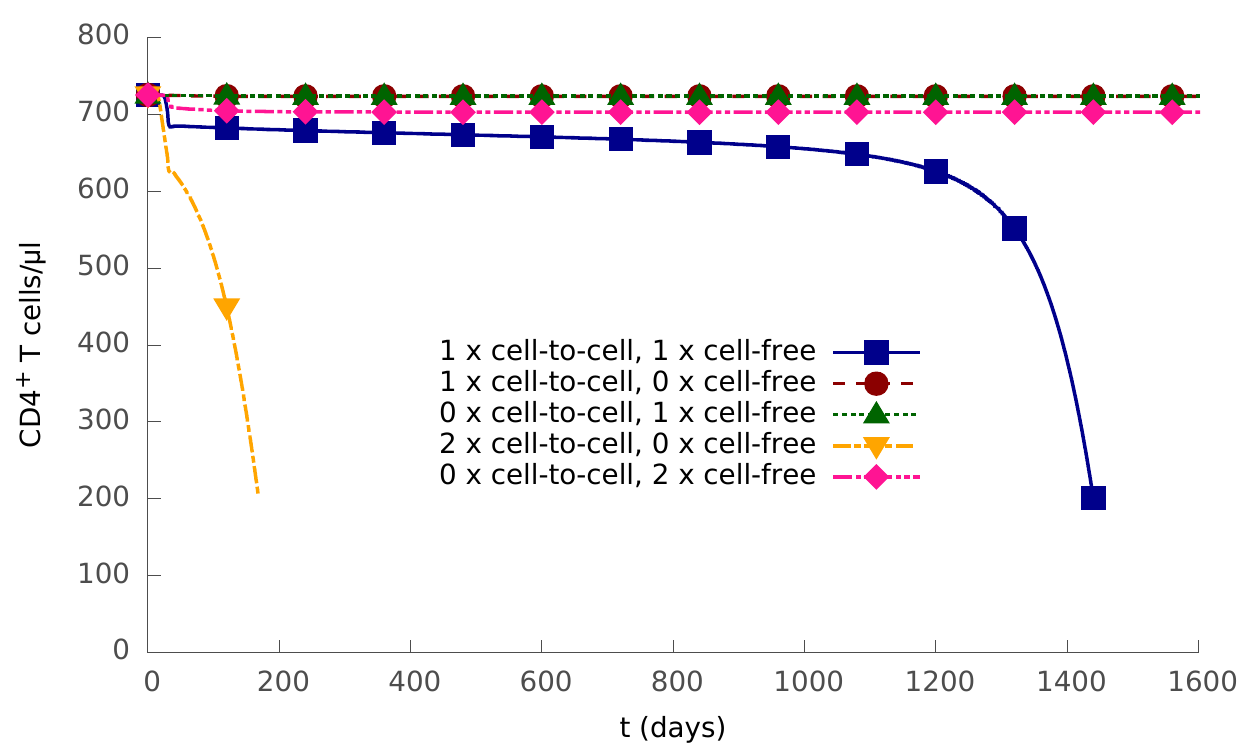}
\end{overpic}
\caption{\label{fig-b1b2}\textbf{Two modes of HIV-1 infection.} The density of CD4$^+$ T cells as a function of time for different values of cell-to-cell infection rate $\beta_1$ and cell-free infection rate $\beta_2$: (1) both use their default value, (2) $\beta_1$ uses its default value and $\beta_2=0$, (3) $\beta_1=0$ and $\beta_2$ uses its  default value, (4) $\beta_1$ is twice its default value and $\beta_2=0$, (5) $\beta_1=0$ and $\beta_2$ is twice its default value.}
\end{figure}

Having confirmed that the model gives realistic estimations and predictions of real clinical data, we investigated the behaviour of the model in more detail. The role of the two spreading routes was further examined by systematic variation of the cell-to-cell infection rate, $\beta_1$, and the cell-free infection rate, $\beta_2$. The predicted outcome of infection are shown in \reffig{fig-b1b2}. When either route is abolished, infection is blocked completely; T cell level returns to normal and virus is cleared after the cellular immune response kicks in. If cell-to-cell spread is removed from the model ($\beta_1=0$) even a doubling in cell-free infection rate does not result in infection progression. In contrast, a doubling of cell-to-cell infection rate increases the set-point of viral load, and greatly speeds up the progression of infection even in the absence of cell-free infectivity. Thus the model suggests cell-to-cell spread may be an important force in allowing virus to establish infection in lymphoid tissue \cite{R26_Parrish_2013}.

\begin{figure}[h]\centering\small
\begin{overpic}[width=0.49\textwidth]{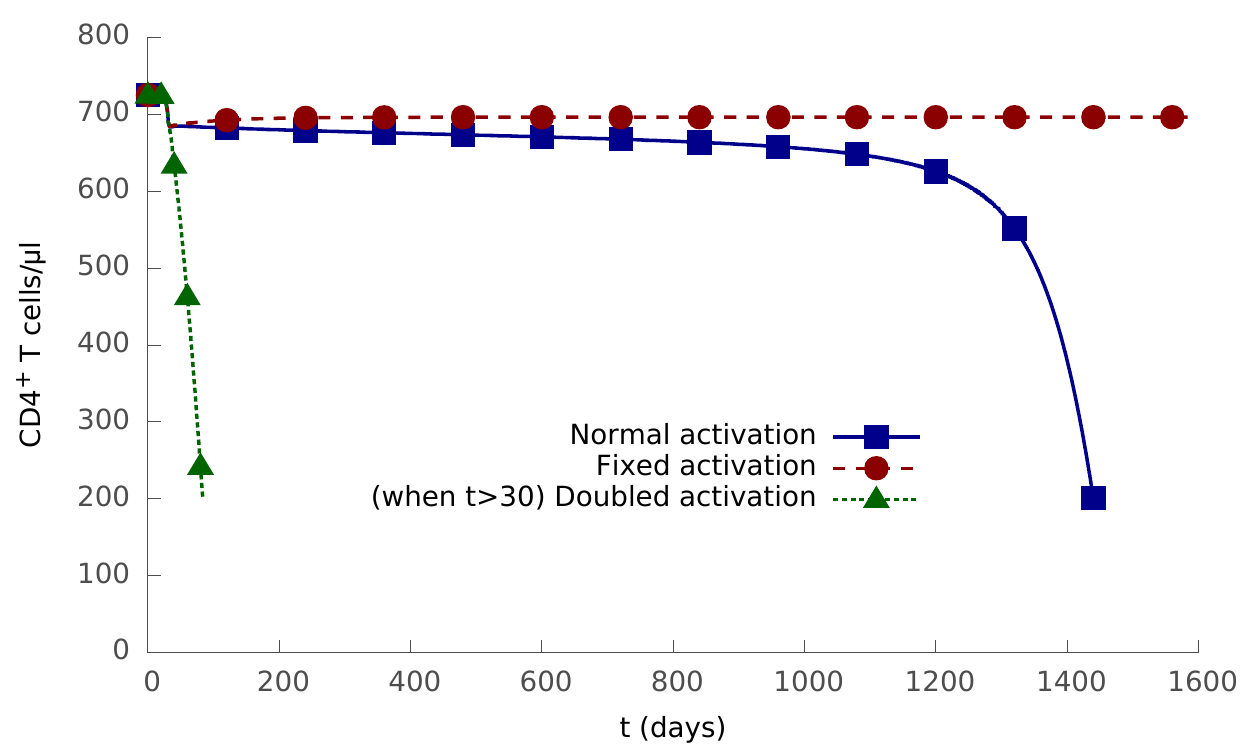}
\put(1,55){\textbf{A}}
\end{overpic}
\begin{overpic}[width=0.49\textwidth]{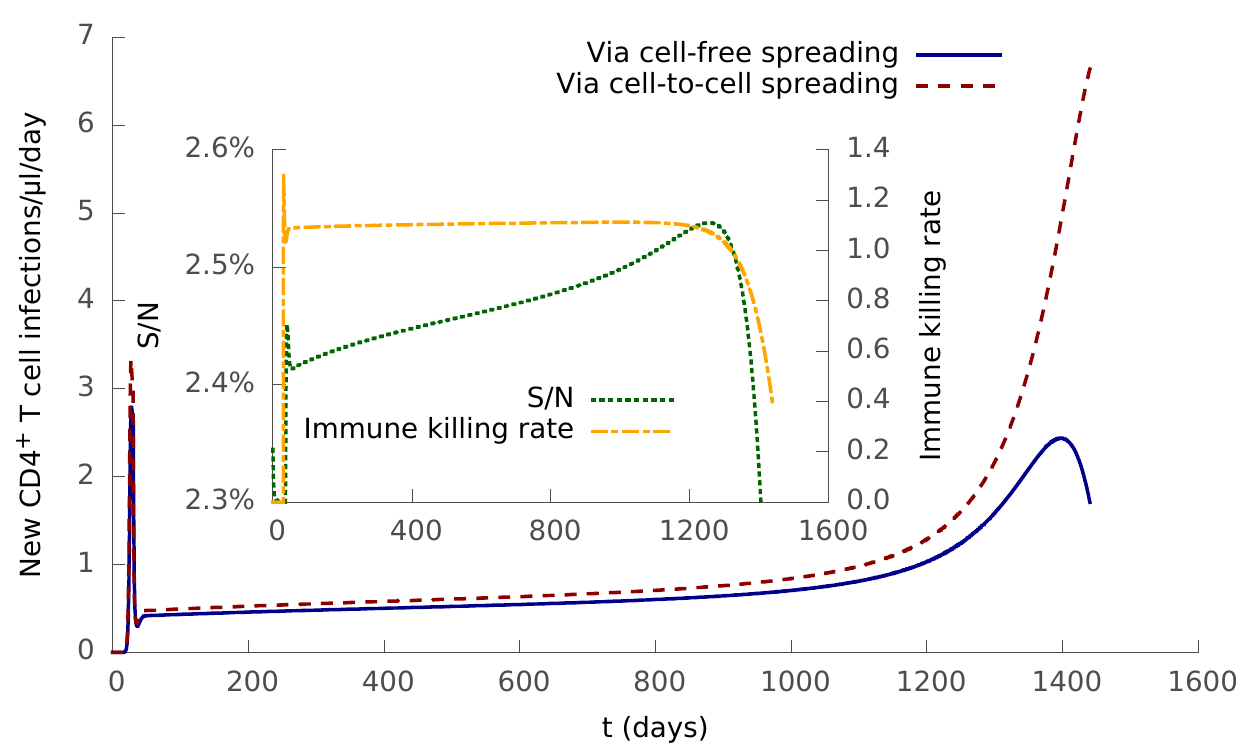}
\put(1,55){\textbf{B}}
\end{overpic}
\caption{\label{fig-adi}\textbf{Progressive CD4$^+$ T cell activation drives progression to AIDS and increased cell-to-cell infection.}\textbf{(A)} Progression of HIV-1 infection for different cell activation rates, including (1) normal activation ($a{N_M}/{N}$ in \refeq{eq:hiv-model2}), (2) fixed activation ($a{N_M}/{N_0}$, where $N_0$ is the initial density of CD4$^+$ T cells), and (3) doubled activation ( $2\times a{N_M}/{N}$ when $t>D$).\textbf{(B)} Numbers of newly infected cells in a day via cell-to-cell spreading and cell-free spreading, respectively. The inset shows  the ratio of susceptible cells to all cells ($S/N$, left y axis) and the strength of immune response ($\kappa \frac{I}{I+0.1} \frac{N}{N_M}$, right y axis) as a function of time, respectively.
}
\end{figure}

In the context of the model, the transition from phase 1 (acute) to phase 2 (stable chronic) is driven by a balance between several processes, including viral spreading through two parallel modes, and the cellular immune response, i.e.\,killing of infected cells as the cytotoxic CD8$^+$ T cell response becomes active. Paradoxically, in the stable chronic phase, the activation of T cells, which is the hallmark of adaptive immunity and is aimed at protecting the host, in fact contributes to the persistence of HIV-1. The role of CD4$^+$ T cell activation is explored in \reffig{fig-adi}A. In this model, the rate of T cell activation $a(N_M/N)$ increases as the number ($N$) of T cells falls, which can be considered to represent a type of homeostatic regulation reinforcing immunological activity relevant to the progressive damage of the immune system and its consequences. In the absence of this feedback (i.e. when activation rate is fixed), HIV infection would not progress to AIDS after the onset of the cellular immune response. In contrast, if the activation rate is doubled, then infection progresses significantly faster to AIDS.  These results confirm and extend the findings of DeBoer and Perelson \cite{R12_De_1998}, which suggested an increasing rate of cellular activation was important in establishment of chronic infection and progression to AIDS. The results are also consistent with evidence that non-pathogenic SIV infection in the natural host species results in viral replication in the absence of chronic immune activation and no AIDS \cite{R28_Silvestri_2007}. \reffig{fig-adi}B depicts the number of CD4$^+$ T cells newly infected via either cell-to-cell spread or cell-free spread as the infection progresses. The model predicts that cell-to-cell transfer becomes increasingly dominant as the total number of CD4$^+$ T cells falls , the proportion of susceptible cells rises (\reffig{fig-adi}B inset left y axis) and the strength of immune response falls (because of immune exhaustion, see \reffig{fig-adi}B inset right y axis).

\subsection*{Treatment evaluation}

Modelling can help evaluate the long term effects of different treatments on HIV-1 progression. Once the start time, duration, and effectiveness against two modes of HIV-1 spread are known for a treatment, its effects on HIV-1 progression can be evaluated by the model.

To validate the model's ability to evaluate HIV-1 treatments, we use it to theoretically reproduce the results of the Short Pulse Anti-Retroviral Therapy at Seroconversion (SPARTAC) trial \cite{SPARTAC_2013}. The SPARTAC trial aims to evaluate how the short-course antiretroviral therapy (ART) delays HIV progression. The patients (366 in total) who participated in the trial were randomly assigned to three groups: standard care, 12-week ART treatment, and 48-week ART treatment. For these three groups of patients, the primary end point $t_p$ (defined as when CD4$^+$ count $\leq 350\,cells/\mu l$ or the start of long-term ART) on average reached 157 weeks (standard care), 184 weeks (12-week ART), and 222 weeks (48-week ART) after randomisation. Randomisation is the time when the trial starts .

We first estimate the trial time points (randomisation, start of the treatments, and primary end points $t_p$) in terms of days after the initial infection (See Materials and Methods). We then use the average CD4$^+$ count and virus load of all patients at randomisation, and the primary end point of the patients in standard care group to fit the five immune-relevant model parameters ($Q_0$, $S_0$, $N_M$, $\kappa$ and $D$). These fitted parameters represent an average patient in the trial.



We then evaluated the effects of 12 and 48 week therapy using the model. We assumed that the therapy was $100\%$ effective against cell-free transmission, but we evaluated the model for both $100\%$ and $50\%$ efficiency against cell-to-cell transmission, since cell-to-cell transmission has been reported as being more resistant to some forms of therapy \cite{Sigal_2011, R37_Titanji_2013}. Both modalities reproduced the observed effects of therapy well, and the model results were robust to changes in the efficiency (\reftab{tab-SPARTAC}). The model therefore not only fits known data sets (standard care group) but also accurately predicts the effects of new treatment regimes on two independent patient groups (12-week and 48-week).

\begin{table}[h]\small\centering
\caption{\label{tab-SPARTAC}\textbf{Model prediction of the SPARTAC trial}}
\begin{tabular}{lr|c|c|c}
\hline
\multicolumn{2}{r}{Patient groups:}  & {Standard-care} & {12-week-treatment} & 48-week-treatment\\
\hline
$N_r$ ($cells/\mu l$) $^*$ &Data: & \multicolumn{3}{c}{$559$} \\
&Model: & \multicolumn{3}{c}{$560$} \\
\hline
$V_r$ ($virons/\mu l$) $^*$ & Data: & \multicolumn{3}{c}{$17$ ~($log_{10}V_r=1.2$)} \\
&Model: & \multicolumn{3}{c}{$12$ ~($log_{10}V_r=1.1$)} \\
\hline
$t_p$ (days)  &Data: & 1,197 & 1,386 & 1,652 \\
&Model: & 1,197 & \textbf{1,392} $|$ 1,394 & \textbf{1,654} $|$ 1,656 \\
\hline
\end{tabular}\\
\begin{flushleft}$N_r$ and $V_r$ are the density of  CD4$^+$ T cells and free virions at the {\it randomisation} (when patients groups are randomly divided and the trial starts). 
$t_p$ is the time between the initial infection and the {\it primary end point} (when density of CD4$^+$ T cells decreases to $350~cells/\mu l$ or the start of long-term ART \cite{SPARTAC_2013}). The $t_p$ estimated from the model is the time when CD4$^+$ T cells decreases to $350~cells/\mu l$. The Materials and Methods section describes how $t_p$ is inferred from the data \cite{SPARTAC_2013} in terms of days after infection.
The model parameters calibrated from the trial data (by fitting the average $N_r$, $V_r$ for all patients and $t_p$ for the standard care group) are 
$Q_0=566~cells/\mu l$, $S_0=13~cells/\mu l$, $N_M=638~cells/\mu l$, $\kappa=1.292165~/day$
  and $D = 28~days$. The trial treatment is assumed to be $100\%$ effective against cell-free HIV-1 spread. We estimate the $t_p$ for 12-week and 48-week treatment groups if the treatment is $50\%$ (value in bold font) or $100\%$ (value in normal font) effective against cell-to-cell spread of HIV-1.

*For the three patient groups, the model predication uses the same $N_r$ and $V_r$, which are the average of all patients in the trial. 
\end{flushleft}
\end{table}

\begin{figure}[h]\centering\small
\begin{overpic}[width=0.49\textwidth]{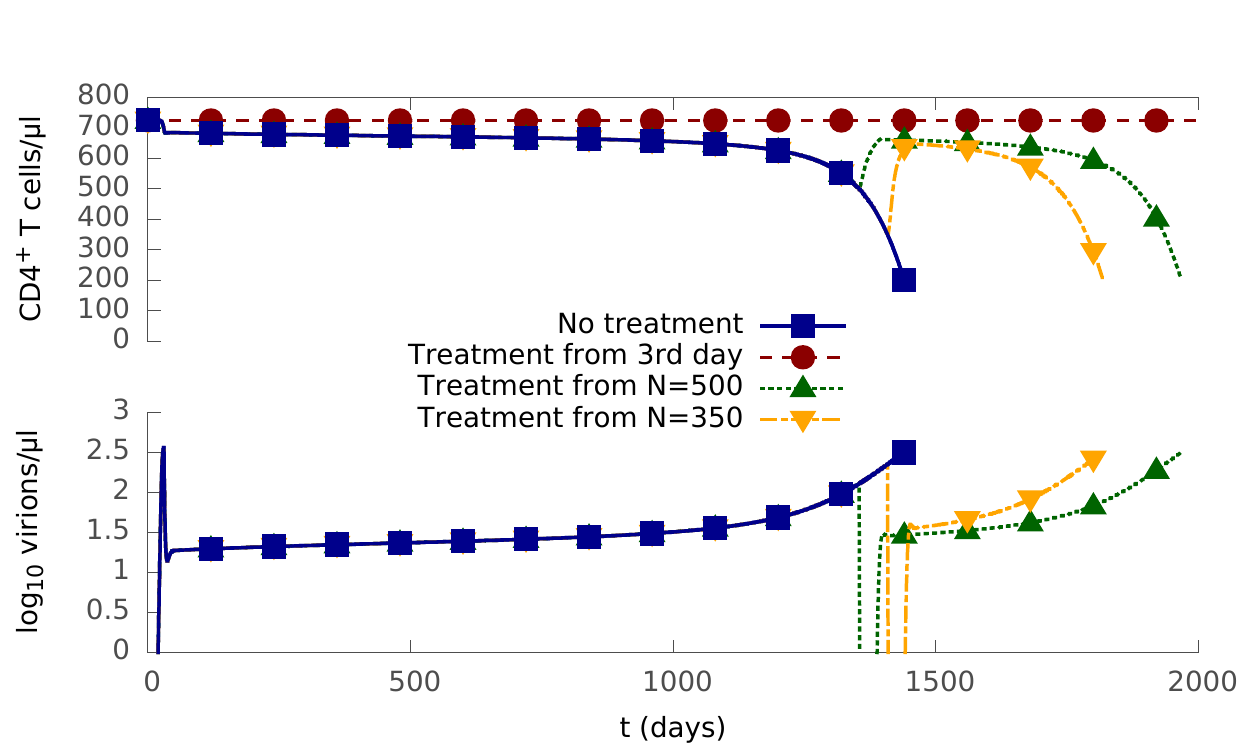}
\end{overpic}
\caption{\label{fig-dtrt}\textbf{Impact of treatment starting time on HIV progression.} HIV progression for a 30-day `perfect' treatment starting at three different times after the initial infection: (1)~on the 3rd day when the density of all CD4$^+$ T cells is $N=725\,cells/\mu l$, (2)~when $N=500\,cells/\mu l$; (3)~when $N=350\,cells/\mu l$. The `prefect' treatment here means both cell-to-cell infection and the cell-free infection are completely blocked (i.e.~$\beta_1=0$, $\beta_2=0$ and the virus release rate $g=0$) for 30 days.}
\end{figure}

We therefore further explored the sensitivity of the model to perturbation as a function of treatment starting time (\reffig{fig-dtrt}). The ``treatment'' lasts for 30 days, during which both cell-free and cell-to-cell infection are completely blocked. Once ``treatment'' is finished, two modes of HIV-1 infection resume. Early treatment in this model (3 days after infection, i.e. post-exposure prophylaxis) leads to no decline in CD4$^+$ T cell density, and no chronic infection phase. The same treatment applied when T cell density reaches the levels ($500$ CD4$^+$ T $cells/\mu l$ and $350$ CD4$^+$ T $cells/\mu l$) at which the World Health Organization recommends ART initiation \cite{R29_WHO_2013} is followed by  a rapid virus rebound after the treatment stops, and the disease progresses according to its normal course. Thus , as HIV-1 progresses it becomes increasingly difficult to control infection in this model.

\begin{figure}[h]\centering\small
\begin{overpic}[width=0.49\textwidth]{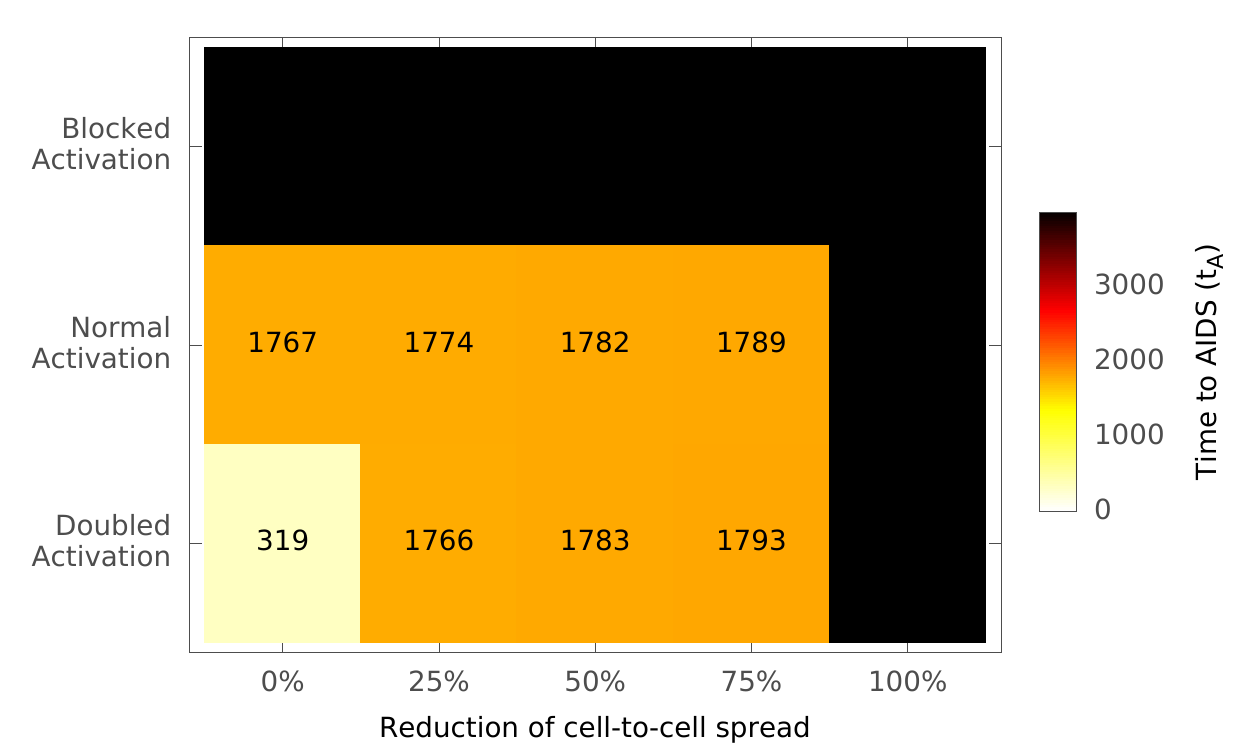}
\put(7,53){\textbf{A}}
\end{overpic}
\begin{overpic}[width=0.49\textwidth]{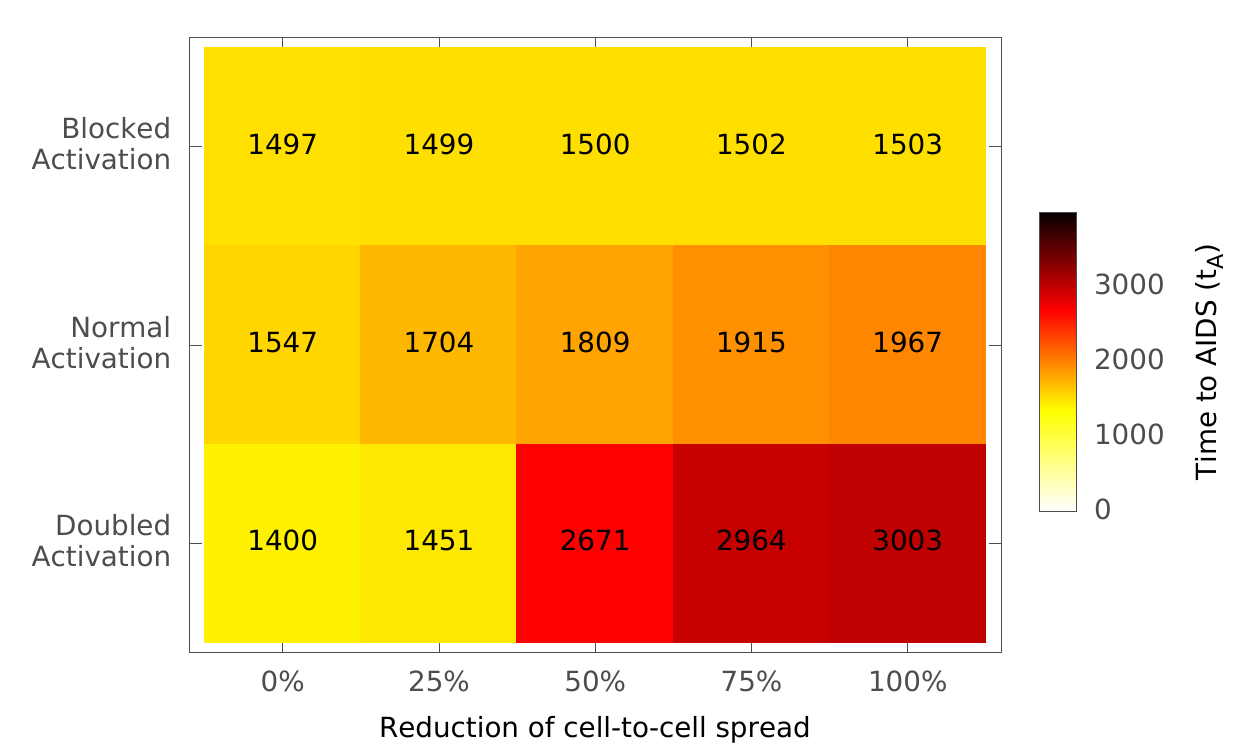}
\put(7,53){\textbf{B}}
\end{overpic}
\caption{\label{fig-tatrt}\textbf{Prediction of the time to AIDS, $t_A$, for different treatment schemes.}We consider 30-day treatments starting \textbf{(A)} on the 3rd day and \textbf{(B)} when $N=500\,cells/\mu l$, respectively. All treatments block cell-free infection completely ($\beta_2=0$ and $g=0$).  The treatments also affect cell-to-cell infection (x axis) and / or manipulate the CD4$^+$ T cell activation process (y axis, see \reffig{fig-adi}A). The black squares represent treatments that reduced the density of infected cells, latent cells and free virions all below $10^{-12}$ per $\mu l$.}
\end{figure}

Finally, we looked at the interactions between treatment starting time, activation rate and efficiency of therapy against cell-to-cell spread (\reffig{fig-tatrt}). In general, increased efficiency of therapy and earlier treatment both prolonged time to progression to AIDS. However, the effects of altered activation depend in a complex way on the context of the intervention. Blocking activation early is beneficial, since it will reduce the number of susceptible cells HIV-1 can infect; while blocking activation late, when the latent HIV-1 reservoir has been established, will prevent latent HIV-1 from being activated and eradicated. Increasing cellular activation, which has been proposed as a means of flushing out the latent reservoir \cite{Wightman_2012}, can be effective in increasing time to AIDS when given in the context of efficient anti-viral therapy, but can shorten the time if concomitant anti-viral therapy is incomplete. This is because increasing cellular activation increases both the number of susceptible cells (activated from quiescent cells) and the number of infected cells (activated from latent cells) at the same time. Thus if it is used together with an effective anti-viral therapy, the latent HIV-1 reservoir will be flushed out and killed by the antiviral drugs. But if the concomitant anti-viral therapy is not efficient enough to clear the increased number of infected cells, the spread of HIV-1 will speed up.

\section*{Discussion}

The aim of this study was to develop a model that reduces the enormous biological complexity of the course of HIV-1 infection to a few well-defined equations but nevertheless retains the main dynamic features of the disease. In particular, the model was required to predict the evolution of a long lived metastable state of low level viral infection, which ultimately breaks down to uncontrolled viral growth and a precipitous fall in CD4$^+$ T cells, two hallmarks of AIDS. The model incorporates both an immune response, which is believed to be a major factor limiting viral expansion by killing of infected cells, and the formation of long lived latently infected cells, which are believed to play an important role in limiting the long-term effects of antiviral therapy. The key distinguishing features of the model are that it incorporates explicitly cell-to-cell spread of virus as well as classical spread via cell-free virus. The motivation of building such a model was to investigate the role of these two modes of spread in determining the outcome of infection, and thus complement the limited \textit{in vivo} experimental and clinical data available addressing this question.


A fundamental prediction of the model is that, given the known experimentally derived parameters of cell-to-cell and cell-free spreading, the two modes of spread complement each other and both make important contributions to disease progression. In addition, it is clear that there is a close relationship between the proportion of activated T cells, cell-to-cell spread and disease progression. Specifically, cell-to-cell spreading is strong when the percentage of activated, and therefore susceptible cells is high in the population, since an infected cell is then more likely to encounter an effective target (a susceptible cell) to infect. When the percentage of susceptible cells is low, infected cells will mostly encounter quiescent/resting cells that will provide ineffective targets. These conditions may occur both early and late in HIV-1 infection. The site of infection itself (for example the vaginal mucosa) may contain large proportion of activated T cells some of which may be interacting with APC, particularly if there is a concomitant sexually transmitted infection. The well-documented association between HIV-1 infection and other mucosal infections may therefore reflect the key importance of a high concentration of APC and activated T cells in early transmission of virus \cite{R30_Ward_2010}. There is also convincing evidence that gut associated lymphoid tissue is a major site of viral replication early on in disease \cite{R31_Brenchley_2004,R33_Veazey_1998}. This tissue is characterized by an unusually large proportion of T cells capable of supporting HIV-1 replication, presumed to result from chronic exposure to the gut microbiome. Since cell-to-cell spread is much more efficient, and under these conditions the number of activated target cells are not limiting, our model predicts that gut lymphoid tissue would provide an ideal microenvironment for rapid propagation of HIV-1, at least until the majority of target cells are infected or die. Further development of the basic model, to allow heterogeneity associated with different anatomical compartments would allow this prediction to be tested directly.

The model also predicts a dominant role for cell-to-cell spread in the late phase of HIV-1. As in the scenario early in infection in the gut lymphoid tissue, our model predicts that late stage disease will be associated with a large number of infected cells, combined with a large proportion of activated target cells. A high number of infected cells is a simple corollary of the very high levels of free virus late in infection in the absence of treatment. There is also substantial experimental evidence for increased immune activation in the late phases of HIV-1 \cite{R2_Miedema_2013}, and indeed this has been proposed as an important contribution to pathogenesis. Thus cell-to-cell spread is likely to become the dominant mode of transmission in the late stages of HIV-1. This may be important in light of recent data showing that the different components of current HAART display variable efficacy against cell-to-cell spread \cite{Sigal_2011, R37_Titanji_2013}.

Interestingly, the hybrid spreading mechanism employed by HIV-1 is reminiscent of those of some computer worms such as Red Code II and Cornficker \cite{Moore_2002,Zhang_2014}, which allocate their resources between probing for susceptible target computers in local area networks and globally across the internet. Similarly to HIV-1, local interactions have a high chance of success but access only a limited number of targets while global spread targets a much larger number of targets with a much lower probability of success. Modelling studies have shown that this hybrid spreading is required to explain the large outbreak of such worms on the Internet \cite{Zhang_hm_2014}. It is tempting to speculate that hybrid spreading may contribute to the pathogenicity and dynamics of infection of other viruses that employ parrallel spreading mechanisms, for example Hepatitis C virus \cite{Zeisel_2013}.

The current model incorporates some simplifying assumptions.  Nevertheless, the model does provide some insights into the effectiveness of therapy at different stages of disease. Specifically, the model strongly supports the hypothesis that interfering with viral infection early in HIV-1 progression is likely to have a major impact on the subsequent progress of the disease. Interfering with viral transmission is predicted by our model to be much more effective early in HIV-1 infection. The clinical decision about when to start therapy remains a matter of debate. Current WHO guidelines suggest commencing treatment at CD4$^+$ T cell density of $>350 cells/\mu l$ and $<500 cells/\mu l$ \cite{R29_WHO_2013}. However, studies exploring much earlier commencements of treatment, have claimed increased efficacy \cite{R36_Bacchus_2013, SPARTAC_2013}. For example, the recent ``Short Pulse Anti-Retroviral Therapy at Seroconversion'' (SPARTAC) trial, demonstrated a long term clinical benefit of a limited period of ART soon after seroconversion \cite{SPARTAC_2013}. Our model accurately predicted the results of the SPARTAC trial providing further support for the model's generality and robustness.

Anti-viral therapies that specifically target cell-to-cell spread are not currently available, but are clearly important therapeutic goals \cite{R37_Titanji_2013}. A number of previous studies have proposed combining antiviral therapies with therapies that either limit CD4+ T cell activation (thus reducing the number of target susceptible cells) or increased T cell activation, thus flushing out reisdual latent cells.  These approaches have not given clear cut clinical benefits \cite{R38_Rizzardi_2002,R39_Lederman_2006}. Indeed our model suggests that the outcome of such manipulation of cellular activation will be critically dependent on the time at which it is administered, and the efficiency of concomitant antiviral therapy. Targeted suppression of CD4$^+$ T cell activation, in combination with antiviral therapies may nevertheless offer a useful approach, if used early on in infection.

Mathematical models provide an important tool for understanding and predicting the course of natural HIV-1 infection that complements clinical studies. The most appropriate therapeutic regimens for HIV therapy continue to remain the subject of disagreement and debate. In addition, many new therapeutic modalities aimed at achieving viral eradication, such as the HDAC inhibitors \cite{Wightman_2012}, or therapeutic vaccination \cite{Lu_2004} are being proposed. However, testing new treatment regimens is a costly and time consuming task, and the logistic challenges and expense of running clinical trials to evaluate and compare treatments remain a major bottleneck to translational advances in HIV therapy. Mathematical models have proved of value in the past, but have suffered from omitting important biological processes, thus compromising their ability to accurately recapitulate clinical observations. Our model explicitly incorporates cell-to-cell transmission, and changing rates of cellular activation, two processes that are known to be a key feature of HIV infection. With increased sophistication, and hence ability to accurately model the known biological drivers of disease progression,  mathematical models can become increasingly important in preclinical testing of modified or novel HIV therapies. The model developed here provides specific predictions which emerge from the close links between CD4$^+$ T cell activation and cell-to-cell spread, and their combined contribution to both early and late phases of HIV-1. These predictions emphasize the potential benefits of early or prophylactic treatment with antiretroviral therapies, and suggest that drugs with the ability to effectively block cell-to-cell spread may provide significant therapeutic benefit in long term management or eradication of HIV-1 infection.

\section*{Materials and Methods}
\subsection*{Recruitment and follow-up of HIV-1 infected patients (London data)}

Individuals acutely-infected with HIV-1 were recruited at the Mortimer Market Centre for Sexual Health and HIV Research (London, UK). Subjects were mostly male Caucasians who presented with symptoms of acute retroviral illness.  Patient viral loads and CD4$^+$ T cell counts were measured longitudinally at serial time-points following infection using standard clinical tests. All subjects were offered anti-retroviral treatment at diagnosis. The subjects selected for inclusion in the  study all chose not to receive anti-retroviral therapy in acute or early infection, and remained untreated until disease progression, evidenced by a substantial decline in their circulating CD4$^+$ T cell count occurred.

\subsection*{Ethics Statement}

Patients provided written informed consent for study participation. Study approval was obtained from The National Health Service Camden and Islington Community Local Research Ethics Committee.

\subsection*{Selection of patients to be included in the study (London data)}

There were 39 patients in total in the data from Mortimer Market Centre for Sexual Health and HIV Research (London, UK). For this study, we focused our analysis only on patients with more than ten data points for both HIV load and CD4$^+$ measurements (29 out of the 39 patients). We also excluded from the analysis a further 12 patients who showed no overall decrease in CD4$^+$ count, or no increase in viral load at later timepoints. The focus of the model described above is to capture the ``typical'' characteristics of HIV infection, which include that CD4$^+$ count falls in general, and viral load increases in general as infection progresses. It is widely accepted that in some patients (for example the so-called ``elite controllers'') viral load remains low or undetectable and CD4$^+$ count remains unchanged for long periods. The mechanisms responsible for these phenomena are still incompletely understood. The current model does not attempt to incorporate any such mechanism, and this group of patients was therefore not included in the study. Further elaboration of the current model to include additional features of viral control will be informative in helping to understand such patients. The identifiers for the remaining 17 patients are: MM1, MM4, MM8, MM9, MM12, MM13, MM23, MM24, MM27, MM33, MM39, MM40, MM42, MM43, MM45, MM57, MM60. 

\subsection*{Estimation of the time points of the SPARTAC trial}

According to \cite{SPARTAC_2013}, the median interval between seroconversion and randomization (start of trial) was 12 weeks. The exact average time of seroconversion for patients in the SPARTAC trial is not directly available from \cite{SPARTAC_2013}. We assume it is 2 weeks after initial infection, as seroconversion normally happens within a few weeks after HIV-1 infection. Then the time of randomization (start of trial) can be estimated as $7(2+12)=98\,days$ after infection. The treatment is estimated to start 3 days after randomization \cite{SPARTAC_2013}, i.e. $98+3=101\,days$ after infection. And the primary end points for patients in standard care, 12-week ART, and 48-week ART groups are respectively: $98+7\times157=1197\,days$, $98+7\times184=1386\,days$, and $98+7\times222=1652\,days$ after infection.

\subsection*{Availability and implementation}

Calculation results in this paper are obtained using LeoTask \cite{LeoTask_2015}, a parallel task running and results aggregation framework that we have developed for computational research. The framework and an executable programme that implements our HIV-1 model are both freely available at \url{http://github.com/mleoking/leotaskapp}.

\section*{Acknowledgments}

We are extremely grateful to the patients who provided samples from which the data modelled in this study was derived.




\bibliographystyle{plos2015_pp}
\bibliography{references}






\clearpage


\section*{Supporting Information}

\subsection*{S1 Table}
\label{tab-fit-pars}\textbf{Model parameters calibrated from a cohort of treatment-naive HIV-1 patients}. Results are shown for all 17 patients. In the data, time is recorded relative to the first appearance of symptoms of HIV infection. As the actual initial infection date is unknown, we assumed a constant ``eclipse'' phase of 20 days between initial infection and first appearance of symptoms.

\subsection*{S2 Table}	
\label{tab-fit-vars}\textbf{Model prediction of a cohort of treatment-naive HIV-1 patients.} $N_S$ is the quasi-steady CD4$^+$ T cell density and $V_S$ is the quasi-steady density of free virions, which are average densities between the 100th and 800th days after initial infection. $t_A$ is the time to AIDS, which is defined as the time between initial infection and when the density of CD4$^+$ T cells falls  to 200 $cells/\mu l$.

\subsection*{S1 Fig}
\label{fig-constn}\textbf{HIV-1 infection course generated by the HIV-1 model using two different cell-to-cell infection terms}: 1) the normal cell-to-cell infection term ($c\theta\beta_1IS/N$) in \refeq{eq:hiv-model2}, and 2) the cell-to-cell infection term that fixes $N$ to be $N_0$ ($c\theta\beta_1IS/N_0$ where $N_0$ is the total number of CD4$^+$ T cells before HIV-1 infection and it is a constant). With a fixed $N=N_0$ in the term, the HIV-1 model would not be able to recapitulate the AIDS phase.

\subsection*{S2 Fig}
\label{fig-vdi}\textbf{$V/I$ (density of free virions divided by the density of infected CD4$^+$ T cells) as a function of time.} There are some obvious changes in $V/I$ during the early days of HIV-1 infection. The figure however also supports some existing works' \cite{Deutekom_Boer_2013} assumption that $V$ and $I$ are proportional to each other during the chronic (stable) phrase of HIV-1 infection. Our model depicts HIV-1 infection in all its three phrases and we do not therefore assume $V$ to be proportional to $I$.

\end{document}